\documentclass[useAMS,usenatbib]{mn2e}

\usepackage{graphicx,epsfig}

\def\gsim{\;\lower4pt\hbox{${\buildrel\displaystyle >\over\sim}$}\;}
\def\lsim{\;\lower4pt\hbox{${\buildrel\displaystyle <\over\sim}$}\;}
\def\grls{\;\lower4pt\hbox{${\buildrel\displaystyle >\over <}$}\;}

\title[On variable protostellar mass accretion rates]
{Variable protostellar mass accretion rates in cloud cores}
\author[Y. Gao and Y.-Q. Lou]
{Yang Gao$^{1}$\thanks{E-mail: gaoyang-00@mail.tsinghua.edu.cn
  (YG);\hskip 1.5cm\hbox{},
  louyq@tsinghua.edu.cn (Y-QL)} and
  Yu-Qing Lou$^{2,\ 3}$\footnotemark[1] \\
$^1$ Centre for Combustion Energy and Department of Thermal Engineering,
 Tsinghua University, Beijing 100084, China\\
$^2$ Department of Physics and Tsinghua Centre for Astrophysics (THCA),
 Tsinghua University, Beijing 100084, China \\
$^{3}$ Tsinghua-NAOC
      joint Research Center for Astrophysics,
      Tsinghua University, Beijing 100084, China \\
   }
\date{Accepted 2016 October 13. Received 2016 October 01;
      in original form 2016 January 28}

\begin{document}

\maketitle

\begin{abstract}
Spherical hydrodynamic models with a polytropic equation
  of state (EoS) of forming protostars are revisited for
  the so-called luminosity conundrum highlighted by observations.
For a molecular cloud (MC) core with such an EoS of a polytropic index
  $\gamma>1$, the central mass accretion rate (MAR) decreases with
  increasing time as a protostar emerges, offering a sensible solution
  to this luminosity problem.
As the MAR decreases, the protostellar luminosity also diminishes,
  making it improper to infer the star formation time by currently
  observed luminosity using an isothermal model.
By observations of radial density profiles and radio continua
  of numerous MC cores evolving towards protostars,
  polytropic dynamic spheres of $\gamma>1$ are also preferred.
\end{abstract}

\begin{keywords}
accretion, accretion discs --- (stars:) brown dwarfs ---
hydrodynamics --- ISM: clouds --- gravitation
--- stars: formation
\end{keywords}

\section{Introduction}

The expansion wave collapse solution (EWCS) of \citet{shu1977} is
  well known
  for low-mass star
  formation\footnote{Formation of brown dwarfs
  \citep[][]{andre2012} and super Jupiter planets
  (Sumi et al. 2011)
  can be also described by gaseous spherical collapse models at
  certain stages (Lou \& Shi 2014).},
  while by extensive observations of actual star-forming molecular
  cloud (MC) cores, the protostellar mass accretion rate (MAR)
  disagrees with Shu's isothermal model of a {\it constant} MAR
  \citep[][]{kenyon1994}.
By observing dust extinction and (sub)millimetre continua,
  it is realized that the radial profiles of MC core
  temperature and density do not fit an isothermal model
  \citep[e.g., Shirley et al. 2002;][]{
  harvey2003b,nielbock2012}.
For non-isothermal MC cores, hydrodynamic
  models with polytropic equations of state (EoSs)
  have been studied \citep{
  fatuzzo2004,lou2006,
  wang2008}
  and the special $\gamma=4/3$ case has also been explored
  \citep[][Lou \& Shi 2016, Li \& Lou 2016, Lou \& Li 2016]{goldreich1980,lou2008}.
A combination of the singular isothermal sphere (SIS) and polytropic
  sphere has also been attempted \citep{curry2000}.
In massive star formation and the description of giant MCs, static
  polytropic spheres with $\gamma<1$ \citep[e.g.,][]{
  mckee2002},
  or even logotrope spheres
 \citep[e.g.,][]{mclaughlin1997}
  were proposed to mimic turbulence effects.
Polytropic sphere models with a range of $\gamma$ can
  broaden applications of the isothermal EWCS,
  especially in accommodating current observations
  of MARs and MC core radial density profiles.

For forming protostars in molecular filaments, we recently
 analyzed dynamic collapses of polytropic cylinders under
 self-gravity with or without magnetic fields (Lou 2015;
 Lou \& Xing 2016; Lou \& Hu 2016a,b).
Such cylindrical collapses with axial uniformity and axisymmetry
 may further break up into segments and clumps along the axis by
 Jeans instability, leading to chains or binaries of collapsed
 objects.

One crucial observational test for star
  formation models is the so-called luminosity problem.
\citet{
kenyon1994} found that the protostellar MAR
  derived from the observed luminosity is too low to accumulate
  $\sim 1 M_{\odot}$ mass in the typical embedded phase of
  $\sim 10^5$~yr in the Taurus-Auriga region, if a
  {\it constant} MAR of the isothermal EWCS is adopted.
This luminosity dilemma has been confirmed and aggravated by
  the {\it Spitzer} legacy project ``From Molecular Cores to
  Planet-forming Disks" or ``Cores to Disks," (i.e. c2d)
  and Gould Belt data \citep[e.g.,][]{evans2009,dunham2013}.
Radiative transfer analyses involving the EWCS
  model, a star-disk system, accretion from clumps, or
  episodic MARs have been performed
\citep{terebey1984,myers1998,young2005,dunham2010,myers2011,offner2011,dunham2012,padoan2014,vorobyov2015} towards solving the luminosity problem.
Here the episodic MARs as suggested by \citet{kenyon1995}
  are supported by theoretical studies via mechanisms of
  gravitational instabilities \citep[][]{boss2002,
  kratter2010},
  magnetically driven bursts \citep[][]{tassis2005}, or
  magneto-rotational instabilities \citep[][]{zhu2009,stamatellos2012};
  and by observations of protostellar accretion bursts
  \citep{jorgensen2015} and D/H ratio of water \citep{owen2015}.
Besides, evolving luminosity has also been studied by
 slightly changing the SIS model \citep{offner2011,dunham2014}
  or including turbulence effects in the mass
  infall of MC cores \citep{padoan2014}.
Alternatively, \cite{lou2006} have emphasized that the
  polytropic gas sphere model gives evolving
  protostellar MARs,
  which would offer a physical solution to this
  luminosity conundrum.


Star formation models are also constrained by the
  radial profiles of density and temperature from
  observations of dust extinction \citep[][]{alves2001}
  and (sub)millimeter radio continuum emissions
  \citep[][]{
  motte2001}.\footnote{A combination of dust extinction,
  submillimeter continuum and molecular line features
  further constrains dynamic polytropic collapse models
  \cite[][]{lougao2011,fu2011}.}
Different radial mass density profiles have been inferred
  by fitting dust radio continuum observations in the
  (sub)millimeter bands with assumed temperature
  profiles \citep[][]{vandertak2000,shirley2002} or by
  fitting dust extinction data without prior assumptions
  of temperature profiles \citep[][]{kandori2005,hung2010}.
Various globally static models have been adopted in fitting
  these observations, e.g., Bonnor-Ebert
  spheres \citep[][]{
  alves2001,kandori2005},
  power-law models \citep[][]{vandertak2000,shirley2002,mueller2002,hung2010,miettinen2013},
  double power laws \citep[][]{beuther2002}, or
  SISs \citep[][]{hogerheijde2000b,shirley2002,harvey2003b,kurono2013}.
For the feasibility of
  dynamic self-similar polytropic sphere models with various
  $\gamma$, we show their mass density and temperature profiles
  and make comparison with previous analyses in this Letter.

\section{Dynamic Polytropic Gas Spheres}

To present physical properties of star-forming MCs
  and to compare them with observations, the general polytropic
  (GP) self-similar protostar formation model
  \citep[][Lou \& Shi 2014]{wang2008,lou2008,cao2009,louhu2010}
  is adopted here without magnetic field.
Hydrodynamic partial differential equations (PDEs)
  of spherical symmetry are:
\begin{equation}
  {{\partial\rho}\over{\partial t}}
  +{1\over{r^2}}{{\partial}\over{\partial r}}(r^2\rho u)=0\ ,
  \qquad\qquad
  {{\partial M}\over{\partial t}}+u{{\partial M}\over{\partial
  r}}=0\ ,
  \label{Equ:mass1}
\end{equation}
\begin{equation}
{{\partial M}\over{\partial r}}=4\pi r^{2}\rho\ ,
  \qquad
  {{\partial u}\over{\partial t}}+u{{\partial u}\over
  {\partial
  r}}=-{1\over{\rho}}{{\partial p}\over {\partial
  r}}-{{GM}\over{r^2}}\ ,\label{Equ:force}
\end{equation}
\begin{equation}
  \left(\frac{\partial}{\partial t}+u\frac{\partial}{\partial r}\right)
  {\rm ln}\left(\frac{p}{\rho^{\gamma}}\right)=0\ ,\label{Equ:entropy}
\end{equation}
  with mass density $\rho$, radial bulk flow velocity $u$,
  thermal gas pressure $p$, and enclosed mass $M$ within
  radius $r$ at time $t$; $G$ is the gravitational constant.
PDEs (\ref{Equ:mass1})
   are equivalent mass conservations, the second of PDE
   (\ref{Equ:force}) is the radial
  momentum conservation and PDE (\ref{Equ:entropy}) is
  the conservation of specific entropy along streamlines
  for the GP EoS in the form of
  \begin{equation}
  p=K(r,\ t)\ \rho^{\gamma}\ ,\label{equ:state}
  \end{equation}
  with $K(r,\ t)$
  being a dynamic coefficient depending on $t$ and $r$.
We now consider the conventional polytropic (CP)
  EoS with $n+\gamma=2$ where $n$ is a self-similar
  scaling index in equation (\ref{equ:radius})
   \citep{lou2006}.
For the CP EoS, $K(r,\ t)$ above
  is a global constant.

For a self-similar transformation,
  all physical variables are expressed as products of
  dimensional scaling functions and dimensionless
  self-similar variables, i.e., the radius $r$, radial
  velocity $u$, particle number density $N$,
  thermal gas pressure $p$, and enclosed mass $M$
  bear the following forms:
\begin{equation}
  r=k^{1/2}t^n x\equiv \bar{r}(t) x\ ,
  \label{equ:radius}
\end{equation}
\begin{equation}
  u=k^{1/2}t^{n-1}v(x)\equiv\bar{u}(t)v(x)\ ,
  \label{equ:varu}
\end{equation}
\begin{equation}
  N=\frac{\alpha(x)}{4\pi\mu m_{\rm H}Gt^2}
  \equiv\bar{N}(t)\alpha(x)\ ,
  \label{equ:varr}
\end{equation}
\begin{equation}
  p=\frac{kt^{2n-4}}{4\pi G}\alpha(x)^{\gamma}
  \equiv\bar{p}(t)\alpha(x)^{\gamma}\ ,\
  \label{equ:varp}
\end{equation}
\begin{equation}
  M=\frac{k^{3/2}t^{3n-2}}{(3n-2)G}m(x)
  \equiv\bar{M}(t)m(x)\ ,
  \label{equ:varm}
\end{equation}
  with $\mu$ and $m_{\rm H}$ being the mean molecular weight
  and the hydrogen mass, $n>2/3$ and $\gamma\neq 4/3$.
For the CP EoS, we have $n+\gamma=2$
  and $n=\gamma=1$ is an isothermal gas.
The dimensionless independent self-similar variable
  $x$ is a combination of $r$ and $t$,
  and $\alpha(x)$, $m(x)$ and $v(x)$ are
  dimensionless reduced mass density, enclosed
  mass and radial velocity, respectively.
By mass conservation (\ref{Equ:mass1}), we derive
  \begin{equation}
  m(x)=\alpha(x) x^2 [nx-v(x)]\ .
  \label{equ:m_v_a_relation}
  \end{equation}
The gas temperature $T$ from the ideal gas law is
  \begin{equation}
  T=\frac{\mu m_{\rm H}\ p}{k_{\rm B}\rho }=
  \frac{\mu m_{\rm H}}{k_B}kt^{2n-2}\alpha(x)^{\gamma-1}
  \equiv\bar{T}(t)\alpha(x)^{\gamma-1}\ ,
  \label{equ:temp}
  \end{equation}
  where $k_{\rm B}$ is the Boltzmann constant.
All barred factors of $t$ are
  of the respective physical dimensions.

There are three analytic and asymptotic solutions for
  CP spheres that are important when comparing our
  hydrodynamic model with observations.\footnote{There
  exists also a Larson-Penston type asymptotic solutions
  in the regime of small $x$ for GP gaseous sphere
  without a central point mass \citep[][]{loushi2014}.}
The first is the singular polytropic sphere (SPS) with
\begin{eqnarray}
v=0\ ,\quad\qquad
 \alpha=\bigg[\frac{n^{2}}{2(2-n)(3n-2)}\bigg]^{-1/n}
 x^{-{2}/{n}}\ ,\nonumber\\
 m=n\bigg[\frac{n^{2}}{2(2-n)(3n-2)}\bigg]^{-1/n} x^{(3n-2)/n}\ ,
\label{Equ:static}
\end{eqnarray}
which is a
 globally static equilibrium solution for
 a CP MC core yet singular as
 $x\rightarrow 0^+$.
The second one is the asymptotic dynamic
 solution for $x\rightarrow +\infty$ with
\begin{eqnarray}
\alpha=Ax^{-{2}/{n}}-3(1-1/n)ABx^{-3/n}\ ,
\qquad\qquad\qquad\
\qquad \nonumber \\
v=Bx^{{(n-1)}/{n}} \qquad\qquad\qquad\qquad\qquad\qquad\qquad\qquad
\qquad \nonumber \\
+ \bigg[\frac{2(2-n)A^{1-n}}{n}-\frac{nA}{(3n-2)}
 +\frac{(1-n)}{n}B^2\bigg]x^{{(n-2)}/{n}}\ ,\label{Equ:infinity}
\end{eqnarray}
where $A$ and $B$ are two integration
 constants \citep[][]{loushi2014}.
The last one is the dynamic asymptotic free-fall solution
 towards the MC core centre as $x\rightarrow 0^{+}$:
\begin{equation}
v=-\bigg[\frac{2m_{0}}{(3n-2)x}\bigg]^{1/2}, \qquad\
\alpha=\bigg[\frac{(3n-2)m_{0}}{2x^{3}}\bigg]^{1/2},
\label{Equ:zero1}
\end{equation}
where integration constant $m_0$ is the reduced
  enclosed mass as $x\rightarrow 0^{+}$, representing
  the dimensionless protostellar mass or MAR [see eq.
  (\ref{equ:massaccretion})].
Invoking central free-fall solution (\ref{Equ:zero1})
  and recalling eqns (\ref{equ:varm}) and
  (\ref{equ:m_v_a_relation}), we derive the
  central protostellar MAR as $x\rightarrow 0^{+}$:
  \begin{equation}
  \dot{M_0}=k^{3/2}t^{3(n-1)}m_0/G\equiv\bar{\dot{M_0}}(t)m_0\
  \label{equ:massaccretion}
  \end{equation}
\citep[see][for details]{lou2006,
 wang2008}.

To estimate the time-dependent dimensional scaling factors
  of variables in eqns.
  (\ref{equ:radius})$-$(\ref{equ:varm}), (\ref{equ:temp})
  and (\ref{equ:massaccretion}), we need empirical
  information for MC cores.
According to \citet{myers2005} and \citet{evans2009},
  a typical radius of a star-forming MC core is
  $\sim 0.01-0.1$~pc ($\sim 10^3-10^4$~AU); and a typical
  mean number density in MC cores is estimated by
  $\sim 10^4-10^5~{\rm cm}^{-3}$.
Noting that the dimensionless parts of both
  radius and number density are around unity
 in relations (\ref{equ:radius}) and (\ref{equ:varr}),
  we may choose the length scale and number density scale as
  \begin{equation}
  \bar{r}= k^{1/2}t^n=4\times 10^3~{\rm AU}\ ,
  \label{equ:lengthscale}
  \end{equation}
  \begin{equation}
  \bar{N}= (4\pi G\mu m_{\rm H}t^2)^{-1}
  =9\times10^4~{\rm cm^{-3}}\ ,
  \label{equ:densityscale}
  \end{equation}
  respectively.
Scales of other physical variables can be expressed
  in the following manner accordingly:
  \begin{equation}
  \bar{u}= k^{1/2}t^{n-1}
  =0.30~{\rm km~s^{-1}}\ ,
  \label{equ:realvelo}
  \end{equation}
  \begin{equation}
  \bar{M}= \frac{k^{3/2}t^{3n-2}}{(3n-2)G}
  =\frac{0.41}{(3n-2)}~M_\odot\ ,
  \label{equ:realmass}
  \end{equation}
  \begin{equation}
  \bar{T}= \mu m_{\rm H}kt^{2n-2}/k_B=21~{\rm K}\ ,
  \label{equ:realtemp}
  \end{equation}
  \begin{equation}
  \bar{\dot{M}_0}= k^{3/2}t^{3(n-1)}/G
  =6.4\times10^{-6}~M_\odot~{\rm yr}^{-1}\ ,
  \label{equ:realmassacc}
  \end{equation}
where the mean molecular weight is $\mu\cong2$
 for H$_2$ MCs
 and $M_\odot$ is the solar mass.
The relevant dynamic time-scale
 $t_d= 0.6\times 10^5~$yr
  is estimated by density scale (\ref{equ:densityscale}).

\section{Decreasing Mass Accretion Rate}



\begin{figure}
\epsfig{figure=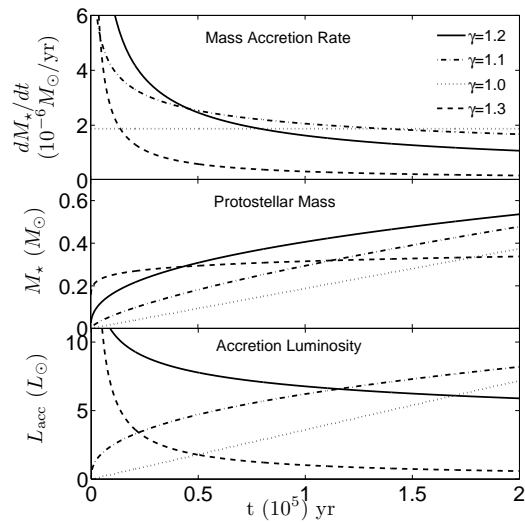,width=8.0cm}
\caption{Time evolution of protostellar MARs
  $d M_*/dt$, masses $M_*$ and accretion luminosity $L_{\rm acc}$
  of dynamic CP MC cores of various $\gamma$ for protostar formation.
For $\gamma>1$ CP MC cores, MARs decrease with increasing time
  (top), and protostars accumulate most of masses
  in their early phases of formation (middle).
In the bottom panel, only for larger
  values of $\gamma=1.2$ and 1.3 do accretion
  luminosity decrease with increasing time.
   \label{fig:rate}}
\end{figure}

Referring to MAR
  scaling (\ref{equ:realmassacc}) at $t_{\rm d}=0.6\times10^5$~yr,
  the protostellar ($r\rightarrow0^+$ for any time $t>0$,
  i.e., $x\rightarrow0^+$) MAR (\ref{equ:massaccretion}) can be
  estimated by $\dot{M_\star}(t)=\epsilon\dot{M_0}(t)$, viz.
  \begin{eqnarray}
  \dot{M_\star}(t)
  =\epsilon\bigg[\frac{t}{0.6\times10^5~{\rm
  yr}}\bigg]^{3(1-\gamma)}6.4\times10^{-6}m_0~M_\odot~{\rm
  yr}^{-1}. \label{equ:protoacc}
  \end{eqnarray}
Here the efficiency coefficient $\epsilon\sim 0.3$, as
 suggested by \citet{alves2007},
 describes the fraction of materials that ends up
  onto a protostar
  \citep[see also][]{evans2009}.
By $t$ integration of eq. (\ref{equ:protoacc}),
  the protostellar mass $M_\star(t)$ becomes
  \begin{eqnarray}
  M_\star(t)
  =\frac{\epsilon t}{4-3\gamma}\bigg[\frac{t}{0.6\times10^5
  ~{\rm yr}}\bigg]^{3(1-\gamma)}6.4\times10^{-6}m_0~M_\odot .
  \label{equ:protomass}
  \end{eqnarray}
Bypassing detailed energy transfer processes, we
  estimate the protostellar accretion luminosity by
  \begin{eqnarray}
  L_{\rm acc}(t)=\frac{GM_\star(t)\dot{M_\star}(t)}{R_\star}
  \qquad\qquad\qquad\qquad\qquad\qquad\nonumber\\
  =\frac{\epsilon^2 t}{(4-3\gamma)}\bigg[\frac{t}{0.6\times10^5~{\rm
  yr}}\bigg]^{6(1-\gamma)}4.2\times10^{-4}m_0^2~L_\odot\ ,
  \label{equ:protolumi}
  \end{eqnarray}
where $L_\odot=3.9\times 10^{33}\ {\rm erg\ s}^{-1}$
  is the solar luminosity, and the protostellar
  radius $R_\star$ is assumed
  at $\sim 3R_\odot$ with $R_\odot=6.9\times 10^{10}\ {\rm cm}$
  being the solar radius \citep[][]{evans2009}.
Realistically, $R_\star$ may vary during a mass accretion
  process and
  $3R_\odot$ here is just a first cut.
For simplicity, we consider four different dynamic
  CP spheres with near-static outer envelopes
  ($B=0$ in eq. (\ref{Equ:infinity}))
  and central free-fall collapsing cores (\ref{Equ:zero1}),
  with dimensionless central mass $m_0=0.975, 1.25, 1.15$ and
  0.26 for $\gamma=1.0, 1.1, 1.2$ and 1.3, respectively
  \citep[][]{gao2009}.
Time evolution of protostellar MAR $\dot{M}_\star$, protostellar
  mass $M_\star$ and accretion luminosity $L_{\rm acc}$ are shown
  in Fig. \ref{fig:rate}.

A solid conclusion of the top panel in Fig. \ref{fig:rate}
  and MAR (\ref{equ:protoacc}) is that CP protostellar
  MARs decrease with increasing time $t$ for $\gamma>1$,
  and in contrast to {\it constant} MARs of isothermal EWCSs
  ($\gamma=1$ and $n=1$), offer a sensible resolution
  to the luminosity problem.
The evolution of protostellar masses (middle panel in Fig.
  \ref{fig:rate}) shows that most protostar masses are
  accumulated in their early phase.
For $\gamma=1.3$, nearly $\sim$ 80\% of the stellar mass is
  accumulated in the first $\sim 10^4$ yrs of formation.
By protostellar accretion luminosity (\ref{equ:protolumi}),
  only for $\gamma>7/6$ dynamic CP spheres do accretion
  luminosity decrease with increasing time (bottom panel
  in Fig. \ref{fig:rate}).
Observations show that bolometric luminosity of
  star-forming MC cores do roughly decrease
  in their early phases (Class 0, I) of star formation
  \citep[fig. 13 in][]{evans2009}, suggesting $\gamma>7/6$
  dynamic CP collapses.

\begin{figure}
\begin{center}
\epsfig{figure=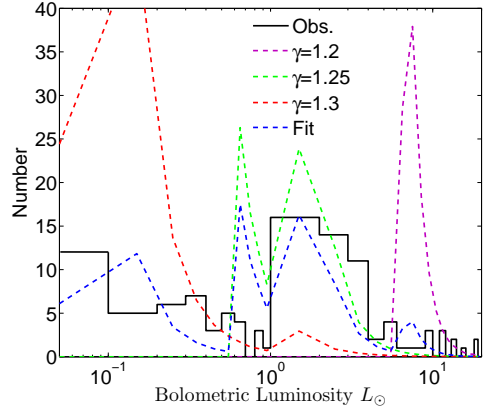,width=8.0cm}
\end{center}
\caption{
PLF or number of protostars with envelopes in luminosity bin
  range $0.1~L_\odot-20~L_\odot$ in a logarithmic scale:
  separate model fits (dash-dot, dashed and dotted curves
  for $\gamma=1.3,\ 1.25$ and $1.2$ CP dynamic core models)
  to c2d PLF observations
  \citep[solid histogram, fig. 14 in][]{evans2009}.
The light solid curve is a combined fit with 65\% of the
  sources having CP EoS of $\gamma=1.25$, and $25\%$ and
  $10\%$ of sources with CP EoS of $\gamma=1.3$ and $1.2$,
  respectively.
The bolometric luminosity is taken as the accretion luminosity
  with a termination limit $L_{\rm ter}=0.08L_{\odot}$.
   \label{fig:histo}}
\end{figure}

With dynamic CP accretion model in hand, we could evolve either
 from core mass function (CMF) for MC cores to
 protostellar mass function (PMF) or from initial mass function
 (IMF) for stars back to PMF;
we follow the latter path.
By setting up the relation between protostellar luminosity
 function (PLF) and PMF, we derive model PLF to compare
 with the observed PLF.
By sampling the dynamic CP model results of Fig. 1, one may fit
 PLF of protostars with envelopes \citep[][]{evans2009,kryukova2012}.
By overall energy conservation,
  the accretion power may be grossly converted to the observed
  bolometric luminosity with $L_{\rm bol}\sim L_{\rm acc}$.
As one cannot follow the luminosity evolution of a single protostar,
  a statistical analysis of a large enough sample would hint at
  evolution track of a single protostar formation.
We assume a continuous and steady start rate of star formation.
  Then the IMF of stars \citep[][]{chabrier2005} is convolved
  \citep{offner2011,myers2011}.
Separate comparisons of our model fits of $\gamma=1.3, 1.25$ and 1.2
  CP spheres with the c2d results \citep[fig. 14 of][]{evans2009} are
  shown in Fig. \ref{fig:histo}.
The intermediate part of the luminosity spectrum is better fitted
  by our CP dynamic sphere of $\gamma=1.25$, while higher and lower
  luminosity parts may involve small contributions from $\gamma=1.2$ and
  $\gamma=1.3$ dynamic spheres, respectively.
The shift of peak sites for different $\gamma$ models
  is due to respective model mean accretion luminosity,
  related to different accretion histories for pertinent $\gamma$
  as shown in Fig. 1.
This fitting suggests that a $\gamma=1.25$ sphere is prominent
  for the early stage of PMF ($\lsim0.2~{\rm Myr}$), with properly
  weighted contributions from other cores of $1.2<\gamma<1.3$.
For illustration, a quick trial
  (light solid in Fig. \ref{fig:histo})
  weighted for the three $\gamma$s leads to a much better overall PLF fit.
Note that the PLF for $L<1L_\odot$ has a width of $0.1L_\odot$
 while for $L>1L_\odot$ the width is $1L_\odot$.
The double peaks in the $\gamma=1.25$ profile and the
 weighted fit are due to this bin width difference.




In addition to protostellar accretion processes in Fig. 1,
  accretion termination by bipolar outflows \citep{shu1988}
  or global expansions \citep[][Fu et al. 2011]{gaolou2010}
  in the late evolution phase for a dynamic CP MC core would
  also affect the PLF.
The observational completeness limit of
  $L_{\rm comp}=0.05-0.10~L_\odot$ plays a role
  in data fit.
The lower luminosity limit $L_{\rm ter}=0.08~L_\odot$
 is adopted.
More realistic fits to the observed PLF require better determination
  of the start rate of star formation, the ending mechanism of mass
  accretion and the CP EoS $\gamma$ distributions.

\section{Radial Mass Density Profiles}

\begin{figure}
\epsfig{figure=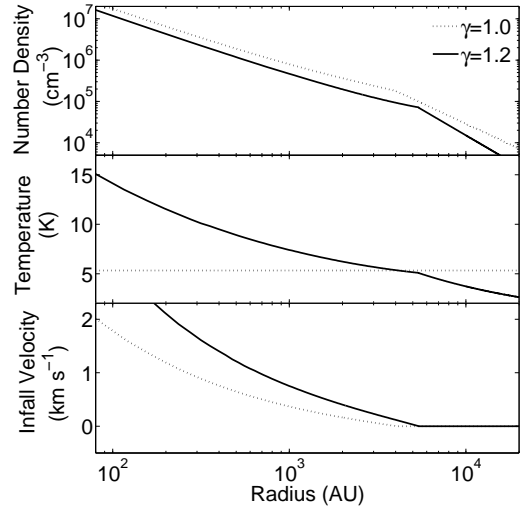,width=8.0cm}
\caption{Radial profiles of number densities (top), gas
  temperatures (middle) and infall velocities (bottom)
  of a CP dynamic MC core.
Different from the isothermal core (dotted curves), a
  CP core of $\gamma=1.2$ has an increasing temperature
  towards the core centre,
  and a steeper decrease in density with increasing radius.
   \label{fig:radial}}
\end{figure}

A successful theoretical model for protostar formation should
  predict radial profiles of density, temperature and velocity
  consistent with observations.
The radial density profile and the temperature profile vary for CP
  spheres with different $\gamma$.
Power-law radial profile ($\rho\propto r^{-p}$) of cloud
  density is an empirical model used to fit observed dust
  continuum distribution.
A power-law index of density distribution achieved by assuming
  certain temperature profiles when fitted to data ranges from
  $p=1$ to $p=2$ \citep[][]{vandertak2000,shirley2002}, the
  boundaries of which are logotropic spheres
 \citep{mclaughlin1997}
  and isothermal spheres \citep{shu1977}, respectively.
However, without prior temperature assumptions, radial density
  profiles inferred from dust extinctions would be more
  realistic \citep[][Hung et al. 2010]{kandori2005},
  and $p\sim 2.5$ for outer envelopes of cloud
  cores is inferred by Hung et al. (2010).

With our CP dynamic models, radial profiles of density,
  temperature and velocity are plotted for
  $t=t_d$
  in Fig. \ref{fig:radial}.
Clearly shown in the top panel of Fig. \ref{fig:radial}, radial
  density profiles of CP spheres with near-static envelopes and
  free-fall collapsing cores are broken into two parts:
  the inner core described by $\rho\propto r^{-1.5}$
  and the outer envelope by $\rho\propto r^{-2/(2-\gamma)}$
  (see eqns (\ref{Equ:infinity}) and (\ref{Equ:static})).
The $\gamma=1.2$ case gives a density profile of
  $\rho\propto r^{-2.5}$ in the outer envelope
  consistent with extinction data in \cite{harvey2003b},
  Hung et al. (2010) and \citet{miettinen2013}.
Double power-law fittings to a large sample of high-mass star
  forming clouds in \citet{beuther2002} also show that a large
  fraction of sources have $p>2$ density indices for MC
  outer envelopes, also hinting at $\gamma>1$ CP spheres.
Fig. \ref{fig:radial} shows that a CP MC core model of $\gamma=1.2$
 naturally provides an increasing temperature towards the centre.
Only two typical profiles of $\gamma=1.0$ and
 $1.2$ are shown in Fig. \ref{fig:radial};
one can show CP models of $1.0<\gamma<4/3$ bearing
 similar profiles but with varying slopes as
 referred to the $\gamma=1.2$ case.
And as a generalization to the velocity field
  of Fig. \ref{fig:radial}, infalling or expanding envelopes
  \citep{lou2006,wang2008} would accommodate more possible
  density and temperature radial profiles than
  near-static envelopes.

\section{Conclusions and Discussion}

The CP dynamic MC core model of forming protostars with
  $\gamma>1$ is advanced as a sensible scenario for
  low mass star-forming MC cores in their early ages, as
(1) it offers a reasonable resolution to the luminosity conundrum
  by allowing a decreasing luminosity with increasing time as
  a protostar dynamically accretes from surrounding envelope
  self-similarly; and
(2) it features density and temperature radial profiles
  more realistic as compared with observations of low
  mass star-forming MC cores.
The empirical fact that the accretion luminosity
  decreases with increasing time corresponds to
  dynamic CP cores with $\gamma >7/6$, also consistent
  with the constraint of averaged radial density profiles.
For such dynamic CP cores of $\gamma>7/6$, most protostellar
  masses are actually accumulated in their early phases.
This would naturally avoid the luminosity problem.

Our CP dynamic MC core model does not necessarily
 exclude other concurrent physical processes.
As suggested recently
 \citep[e.g.,][]{kryukova2012,vorobyov2015}, the
 actual MAR history of an emerging protostar
 may vary for low and high mass stars and is
 most likely a combination of the decreasing
 MAR due to a CP
 EoS and bursty accretions arising from, e.g., disc
 instabilities and shocks (e.g., Shen \& Lou 2004, 2006).
The latter may be more important in late stages of star
 formation after the appearance of a disc, which might also
 be relevant to the low luminosity fit \citep[][]{evans2009}.
Also, either global or bipolar outflows in late stages of
 star formation provides a possible means of accretion
 termination.
As the CP EoS is more adaptable for low mass star formation,
 the corresponding decreasing trend of MAR should be more
 important in the formation history of low mass stars,
 consistent with the results of \citet{padoan2014}.
Further studies connecting CMF, PMF, PLF, IMF, radiation
  transfer and consequences of discs are desirable.

\section*{Acknowledgments}

This research was supported in parts by
MOST grant 2012CB821800, Tsinghua
 CCE and THCA,
 by
 NSFC grants
 10373009, 10533020, 11073014, 11473018
 and 51206088,
 by MOST grant
 2014YQ030975,
 by
 THISRP (20111081008),
 and by the Yangtze Endowment, the SRFDP
 20050003088, 200800030071 and 20110002110008
 as well as 985 grants from MoE
 and AMD scholarships at Tsinghua Univ.
YQL acknowledges support of the China-Chile
 Scholarly Exchange Program
 administered by CASSACA.

\end{document}